\documentclass{aa}
\usepackage{graphicx}
\begin{document}
   \title{The braking indices in pulsar emission models}

   \subtitle{}

   \author{Fei Wu\inst{1,2}, R.X. Xu\inst{2}, Janusz Gil\inst{3}}

   \offprints{R.X. Xu,\\ e-mail: {\tt rxxu@bac.pku.edu.cn}}

   \institute{$^1$National Astronomical Observatories, Chinese Academy of
        Sciences, 20 Datun Road, Chaoyang, Beijing 100012, China\\
        $^2$School of Physics, Peking University, Beijing 100871, China\\
        $^3$Institute of Astronomy, University of Zielona
        G\'ora, Lubuska 2, 65-265 Zielona G\'ora, Poland
        }

   \date{Received~~~~~~~~~~~~~~~~~~~~~; ~~~Accepted}

   \abstract{
Using the method proposed in a previous paper, we calculate pulsar
braking indices in the models with torque contributions from both
inner and outer accelerating regions, assuming that the
interaction between them is negligible. We suggest that it is
likely that the inverse Compton scattering induced polar vacuum
gap and the outer gap coexist in the pulsar magnetosphere. We
include the new near threshold vacuum gap models with
curvature-radiation and inverse Compton scattering induced
cascades, respectively; and find that these models can well
reproduce the measured values of the braking indices.


   }

   \maketitle
%

\section{Introduction}

Why do pulsars spin down? Although it is generally accepted that
pulsars brake due to magnetodipole radiation, this simple model is
questioned by the fact that the observed braking index
$n\equiv\Omega\ddot{\Omega}/\dot{\Omega}^2$ ($\Omega$ is the
angular velocity of rotation) is smaller than 3. For the Vela
pulsar, the index is $n=1.4$, which is the smallest one observed.

Actually, some efforts have appeared to find unusual mechanisms to
understand the observed braking index.
It was suggested previously that the observation of $n\neq 3$ may
result from a force-free precession of a distorted neutron star
(Macy 1974) or the existence of a companion star (Deminanski \&
Proszynski 1979).
Peng et al. (1982) and Huang et al. (1982) suggested that neutrino
and photon radiation from superfluid neutron vortexes may
contribute to pulsar spindown. However, this idea may have
difficulty in interpreting $n<3$ since this mechanism dominates
for older pulsars with longer periods.
Blandford \& Romani (1988) interpreted the observed index by a
multipole field and/or field evolution. Melatos (1997) accounted
for the spin-down of three pulsars (the Crab, PSR B0540-69, and
PSR B1509-58) in the light of a non-standard vacuum dipole model,
where a ``vacuum radius'' is introduced phenomenologically.
Recently, accretion torque has been suggested to explain the
discrepancy (Morley 1993, Menou, Perna \& Hernquist 2001).
Alpar et al. (2002) proposed that both the usual magnetodipole
radiation and the propeller torque applied by the debris disk
formed soon after supernova explosion should cause a pulsar to
spin down (see also, e.g., Marsden et al. 2001), although no
direct evidence exists to show that a normal pulsar has
significant torque of this kind.

An alternative effort, within the framework of ``{\em standard}''
neutron stars and their magnetospheric emission models, was
proposed by Xu \& Qiao (2001; hereafter Paper I).
They addressed that the magnetodipole radiation and the unipolar
generator are two different sources powering a pulsar, and these
two should be combined in studying pulsar spindown. They find that
the calculated braking indices, although model-dependent, are less
than 3, and that a proposed model should be ruled out if it cannot
result in an index as small as 1.4.

In addition, the pulsar emission mechanism is still a great
challenge even over 30 years after discovery.
It is a common point that two classes of accelerators (i.e., the
inner gaps and the outer gaps) may work on pulsar magnetosphere to
reproduce its photon (radio to X-ray or $\gamma-$ray bands)
emission, and two more subclasses of models, i.e., the
space-charge-limited flow models (e.g., Arons \& Scharlemenn 1979;
Harding \& Muslimov 1998) and the vacuum gap models (e.g.,
Ruderman \& Sutherland 1975) for the inner accelerators appear in
the literature.
However, it is quite possible that both inner and outer gaps may
coexist in a magnetosphere (Usov 2000; Paper I) since those two
gaps may work in different field line regions in order to close
the global electric current (e.g., Holloway 1975).

In this paper, using the method developed in Paper I, we calculate
the braking indices in the models where both inner and outer
accelerators lose the rotation energy, neglecting the possible
interaction between those two kinds of gaps.
Since a new polar gap model, the near threshold vacuum gap (NTVG;
Gil \& Mitra 2001) model, became available after Paper I was
published, we will begin with computing the braking indices in
this new model.
It is found that the NTVG model can reproduce an index as small as
1.4, and this model is therefore possible to explain the 5 braking
indices observed.
However, if the inner and the outer gaps coexist, we suggest that
the ICS-induced vacuum gap and the outer gap may work together at
the same time, since the calculated braking indices in this case
are more reasonable than that in other cases.


\section{Braking indices for the NTVG model}

The model of Ruderman \& Sutherland (1975) is ``user friendly'',
but has two imperfect points: 1, the binding energy problem of
ions on the neutron star surface; 2, only half of the neutron
stars are applicable (e.g., Xu, Qiao \& Zhang 1999). Although
these points can be overcome if some radio pulsars are bare
strange stars (Xu et al 1999), Gil \& Mitra (2001) argued that the
RS-type vacuum gaps could also exist for neutron stars with
multipolar surface magnetic fields (i.e., the actual surface
strength is much higher than the dipolar surface component, and
the radius of field curvature is much smaller than the neutron
star radius), since the ion cohesive energy becomes larger if the
field is stronger.

However, in a superstrong surface field, $B_{\rm s}>0.1B_{\rm
q}\simeq 4.4\times 10^{12}$ G, the $\gamma$ photons should be
converted into $e^\pm$ at or near the kinematic threshold (Gil \&
Mitra, 2001); the vacuum gap formed under this condition is thus
called as the Near Threshold Vacuum Gap (NTVG) model.
Gil \& Melikidze (2002) present the gap heights and the gap
potential drops for different pair-production mechanisms (CR-NTVG:
curvature-radiation induced NTVG; ICS-NTVG: inverse-Compton-
scattering induced NTVG) and different estimates of the cohesive
energies of surface iron ions (i.e., Abrahams \& Shapiro 1991,
hereafter AS91; and Jones 1986, hereafter J86).
In this section, we calculate the braking index for the NTVG model
in the frame proposed in Paper I.

For pulsars with given pulsar radius $R$ and polar magnetic field
at surface $B$, the braking index $n$ can be written as
\begin{equation}
   n=3+\frac{\Omega\dot{\eta}}{\dot{\Omega}\eta}=
   3+\frac{\Omega}{\eta}\frac{d\eta}{d\Omega},%
\label{n}
\end{equation}
with
   $$
   \eta\simeq\sin^2\alpha+5.4\times10^{-9}R_6^{-3}B_{12}^{-1}
   \cos^2\alpha\Omega^{-2}\Delta\phi,
   $$
where $\alpha$ is the inclination angle between the magnetic axis
and the rotating axis, $R_6=R/(10^6\textrm{cm})$,
$B_{12}=B/(10^{12}\textrm{G})$, and $\Delta\phi$ is the gap
potential drop which is model dependent.


For the curvature-induced NTVG model, the gap potential drop is
(Gil \& Melikidze 2002)
\begin{equation}
\Delta\phi_{\rm CR}^{\rm NTVG}=4.0\times10^9\zeta^{1/7}
\rho_6^{4/7}b^{1/7}P^{-1/14}\dot{P}_{-15}^{1/14}\ \textrm{cgse},%
\end{equation}
with the parameter $b$ in the form
$$
b_{\rm min}^{\rm CR}=\beta\zeta^{0.08}k^{0.57}\rho_6^{0.32}
P^{-1.15}\dot{P}_{-15}^{-0.5},
$$
where $\beta=52$ for AS91 case and $\beta=1990$ for J86 case,
$\zeta\sim0.85$ is the typical value of the correction factor, $k$
is the heat flow coefficient, $P=2\pi/\Omega$ is the pulsar period
in seconds, and $\dot{P}_{-15}=\dot{P}/10^{-15}$.
For simplicity, one may take $k=1.0$ and $\rho_6=1$, and therefore
has
%
\begin{equation}
   \eta_{\rm CR}^{\rm NTVG}\simeq\sin^2\alpha+13.7\beta^{0.14}R_6^{-2.38}
   B_{12}^{-1}\cos^2\alpha\Omega^{-1.76}.%
\label{cr}
\end{equation}


For the inverse-Compton-scattering-induced NTVG model, the gap
potential drop is (Gil \& Melikidze 2002)
\begin{equation}
   \Delta\phi_{\rm ICS}^{\rm NTVG}=1.7\times10^{10}\zeta^{0.72}k^{-0.14}
   \rho_6^{1.14}b^{-1}P^{-1.22}\dot{P}_{-15}^{-0.5}\ \textrm{cgse},%
\end{equation}
with
   $$
   b_{\rm min}^{\rm ICS}=\gamma\zeta^{0.25}k^{0.34}\rho_6^{0.39}P{-1.1}
   \dot{P}_{-15}^{-0.5},
   $$
where $\gamma=14$ for AS91 case and $\gamma=130$ for J86 case. One
thus has
%
\begin{equation}
   \eta_{\rm ICS}^{\rm NTVG}\simeq\sin^2\alpha+69.6\gamma^{-1}R_6^{2.25}
   B_{12}^{-1}\cos^2\alpha\Omega^{-1.88}.%
\label{ics}
\end{equation}

According to eq.(\ref{n}), (\ref{cr}) and (\ref{ics}), we have
calculated the braking indices in different NTVG models, as
functions of rotation period for typical pulsars with $R_6 = 1$
and $B_{12} = 10$. The results are shown in Fig.1.
It is obvious that the braking index could be smaller than 3 for
any inclination angle in the NTVG model.
We find all these models can have a braking index as small as 1.4.


\section{Braking indices in emission models where the inner and
         outer gaps coexist}

We simply consider the energy conservation between the spindown
power and the ``dipole radiation + inner gap + outer gap'' energy
losses, but ignore the interaction between these two kinds of
gaps. We expect that this simplified case will still be useful
when more detailed and practical consideration is possible in the
future.
First, however, we correct the calculation of braking index\footnote{%
Xu \& Qiao (2001) made a mistake in calculating the braking index
for the outer gap, by letting the gap fractional size $f=1$ when
writing the particle current.
} %
for the outer gap (Cheng Ho \& Ruderman 1986; Zhang \& Cheng 1997)
in Paper I, then we present the results calculated for models
where the inner and outer gaps coexist.

For a self-sustaining outer gap, which is limited by the $e^\pm$
pair produced by collisions between high-energy photons from the
gap and soft X-rays resulting from the surface heating by the
backflowing primary $e^\pm$ pairs, the potential drop is (Zhang \&
Cheng 1997)
\begin{equation}
   \Delta\phi^{\rm OG}=1.6\times10^{12 }R^3_6 B^{-1/7}_{12}
   \Omega^{-10/21}\,\textrm{cgse},
\end{equation}
and the electric current through the gap
\begin{equation}
   F=8.68\times10^{20} B^{3/7}_{12}\Omega^{16/21}\,\textrm{cgse},
\end{equation}
thus the energy loss rate due to the outer gap could be written as
\begin{equation}
   \dot{E}^{\rm OG}=-\Delta\phi^{\rm OG} F=-1.39\times10^{33} R^3_6 B^{2/7}_{12}
   \Omega^{2/7}\,\textrm{erg}.
\end{equation}
In the case where the outer gap engine dominates (i.e., the
inclination angle $\alpha=0$, and only the outer gap works),
$\dot{\Omega}\sim \dot{E}^{\rm OG}/\Omega\sim \Omega^{-5/7}$; we
have therefore a braking index $n=-5/7$ for the outer gap.
According to energy conservation, we obtain the $\eta$ value in a
model where inner and outer gaps coexist,
\begin{equation}
\begin{array}{lll}
   \eta^{\rm IG+OG} & = &
   \sin^2\alpha+(5.4\times10^{-9}R^{-3}_6B^{-1}_{12}\Omega^{-2}\Delta\phi^{\rm IG}\\
   & &
   +2.25\times10^5R^{-3}_6B^{-12/7}_{12}\Omega^{-26/7})\cos^2\alpha,
\label{eta-og}
\end{array}
\end{equation}
where $\Delta\phi^{\rm IG}$ is the potential drop of the inner
gap.
We calculate the braking index, as a function of $P$, from
eq.(\ref{n}) and eq.(\ref{eta-og}).
In order to present the correction to Paper I, the new braking
index calculated for the case where only the outer gap works is
shown in Fig.2.

In the following subsections, we calculate the braking indices in
the inner-outer gap coexisting models.
Three kinds of inner gaps are considered, i.e., the normal vacuum
gap, the near threshold vacuum gap, and the space-charge-limited
flow.

\subsection{Normal VG gap + outer gap}

Various inner accelerator models have been investigated by Zhang,
Harding \& Muslimov (2000).
Here we present only the final results of the $\eta$ parameters
for the related models. Details of the gap potentials can be found
in Xu \& Qiao (2001) and references therein. For the CR-induced VG
case,
\begin{equation}
\begin{array}{lll}
   \eta^{\rm VG+OG}_{\rm CR} & = &
   \sin^2\alpha+(4.97\times10^2\Omega^{-15/7}\\
  & & +2.25\times10^5\Omega^{-26/7})\cos^2\alpha.
\end{array}
\end{equation}
For the ICS-induced VG case,
\begin{equation}
\begin{array}{lll}
   \eta^{\rm VG+OG}_{\rm ICS} & = &
   \sin^2\alpha+(1.03\times10^5\Omega^{-13/7}\\
  & & +2.25\times10^5\Omega^{-26/7})\cos^2\alpha.
\end{array}
\end{equation}

\subsection{NTVG gap + outer gap}
For the CR-induced NTVG case,
\begin{equation}
\begin{array}{lll}
   \eta^{\rm NTVG+OG}_{\rm CR} & = &
   \sin^2\alpha+(1.37\beta^{0.14}\Omega^{-1.76}\\
  & & +2.25\times10^5\Omega^{-26/7})\cos^2\alpha.
\end{array}
\end{equation}
For the ICS-induced NTVG case,
\begin{equation}
\begin{array}{lll}
   \eta^{\rm NTVG+OG}_{\rm ICS} & = &
   \sin^2\alpha+(6.96\gamma^{-1}\times10^2\Omega^{-1.88}\\
  & & +2.25\times10^5\Omega^{-26/7})\cos^2\alpha.
\end{array}
\end{equation}

\subsection{SCLF + outer gap}
For the CR-induced SCLF case,
\begin{equation}
\begin{array}{lll}
   \eta^{\rm SCLF+OG}_{\rm CR} & = & \sin^2\alpha+(38.3\Omega^{-7/4}\\
  & & +2.25\times10^5\Omega^{-26/7})\cos^2\alpha.
\end{array}
\end{equation}
For the ICS-induced SCLF case,
\begin{equation}
\begin{array}{lll}
   \eta^{\rm SCLF+OG}_{\rm ICS} & = & \sin^2\alpha+(2.27\Omega^{-8/13}\\
  & & +2.25\times10^5\Omega^{-26/7})\cos^2\alpha.
\end{array}
\end{equation}

The corresponding braking indices can be calculated from
eq.(\ref{n}) and one of eq.(10)-eq.(15).
Fig.3 shows the braking index curves, as functions of pulsar
periods. It seems that, except for the VG(ICS)+OG model, the
energy loss of the outer gap dominates.


\section{Conclusion \& discussion}

The braking index in different NTVG models are calculated.
The minimum braking indices, $n_{\rm CR}^{\rm
NTVG}(\alpha=0^\circ) = 1.24$ and $n_{\rm ICS}^{\rm
NTVG}(\alpha=0^\circ) = 1.12$, are smaller than 1.4, the smallest
one observed.
This indicates that the NTVG model passes the test proposed by Xu
\& Qiao (2001), and that this model may work for radio pulsars.
In the calculation, typical parameters, e.g.,
$k=\rho_6=R_6=B_{12}=1$ (except $B_{12}=10$ in NTVG models), are
employed. Nonetheless, we find that the general result does not
change significantly if these parameters are adjusted reasonably.

It is suggested that the ``ICS-VG + Outer'' gaps prefer to work in
the pulsar magnetospheres, since other inner-outer gap models
imply very small (even a negative) value of the braking index,
while the observed indices are in the range from 1.4 to 2.9.
These other models may work if much smaller braking indices are
observed.
For NTVG models, there is little difference between the CR-induced
and the ICS-induced ones, because the energy loss rate from NTVG
is relatively small compared with that of the outer gap.

\begin{acknowledgements}
This work is supported by National Nature Sciences Foundation of
China (10273001), by the Special Funds for Major State Basic
Research Projects of China (G2000077602), by Chinese Undergraduate
Research Endowment in Peking University, and by the Grant
2~P03D~008~19 of the Polish State Committee for Scientific
Research. JG would like to acknowledge the great hospitality of
the Department of Astronomy of Peking University, where this work
was started during his visit. RXX thanks Prof. G.J. Qiao for his
valuable general discussions about pulsars.
\end{acknowledgements}


\clearpage

\begin{figure}[t]
  \centering
  \begin{minipage}[t]{.5\textwidth}
    \centering
    \includegraphics[width=7cm]{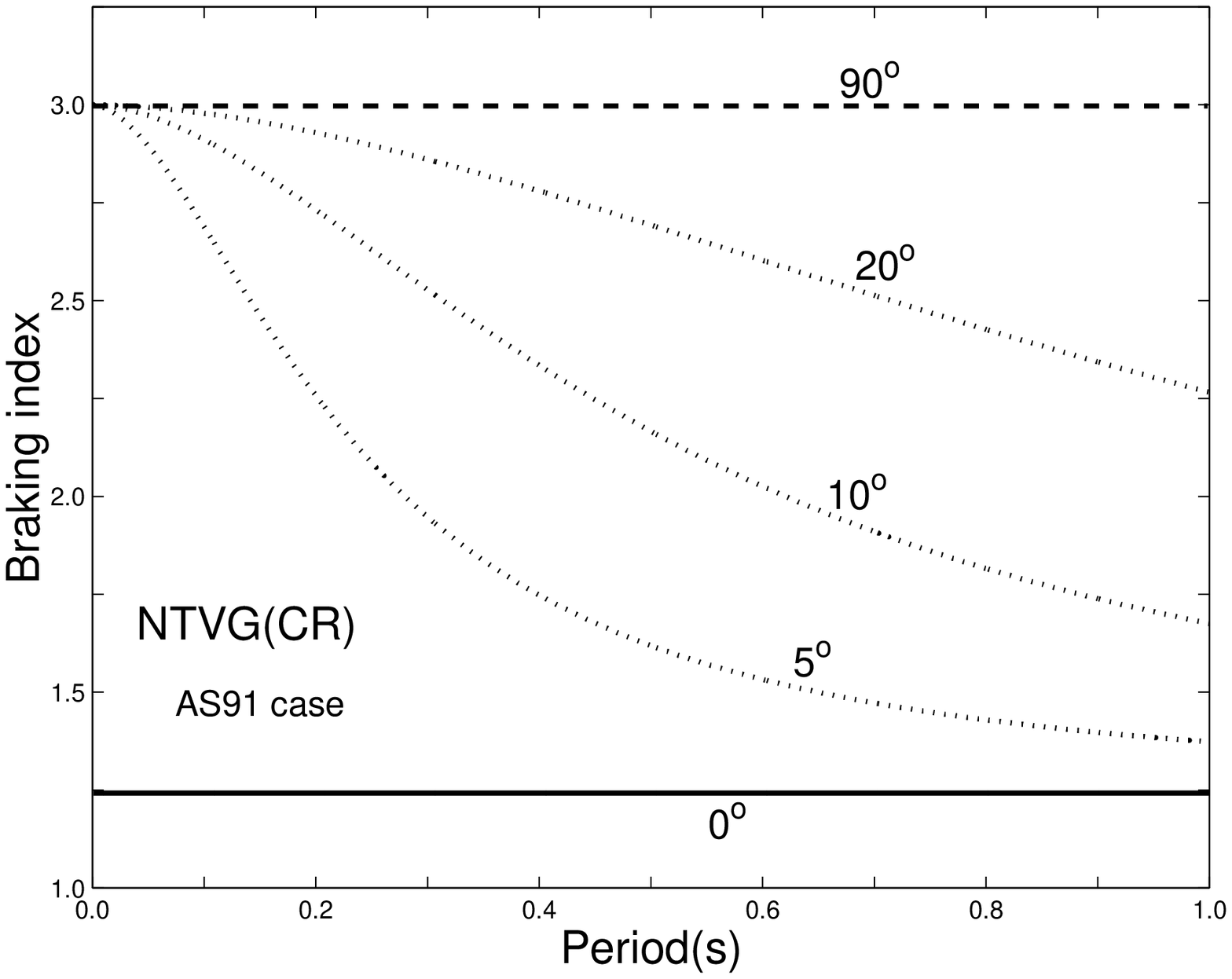}
  \end{minipage}%
  \begin{minipage}[t]{.5\textwidth}
    \centering
    \includegraphics[width=7cm]{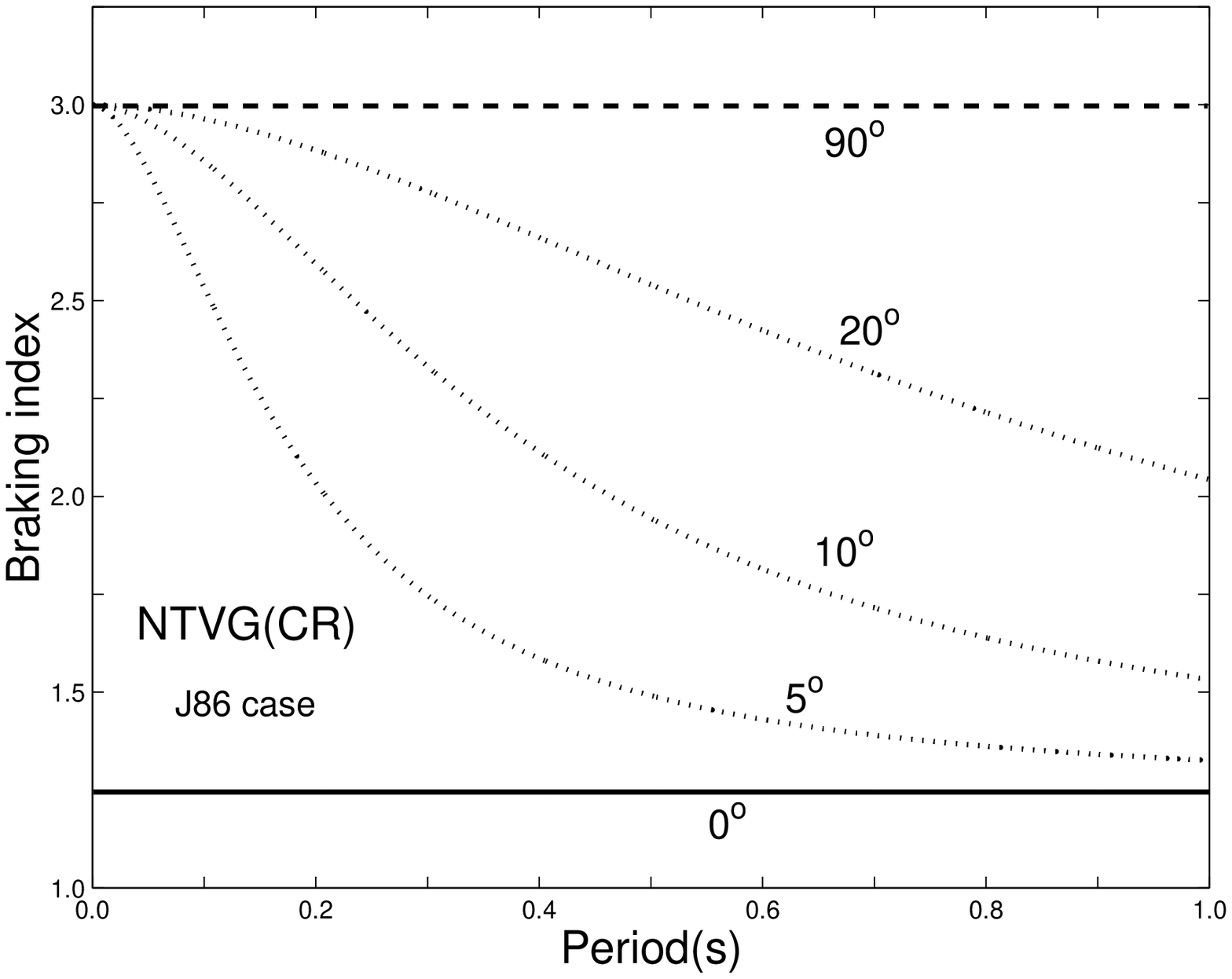}
  \end{minipage}%
\end{figure}

\begin{figure}
  \centering
  \begin{minipage}[t]{.5\textwidth}
    \centering
    \includegraphics[width=7cm]{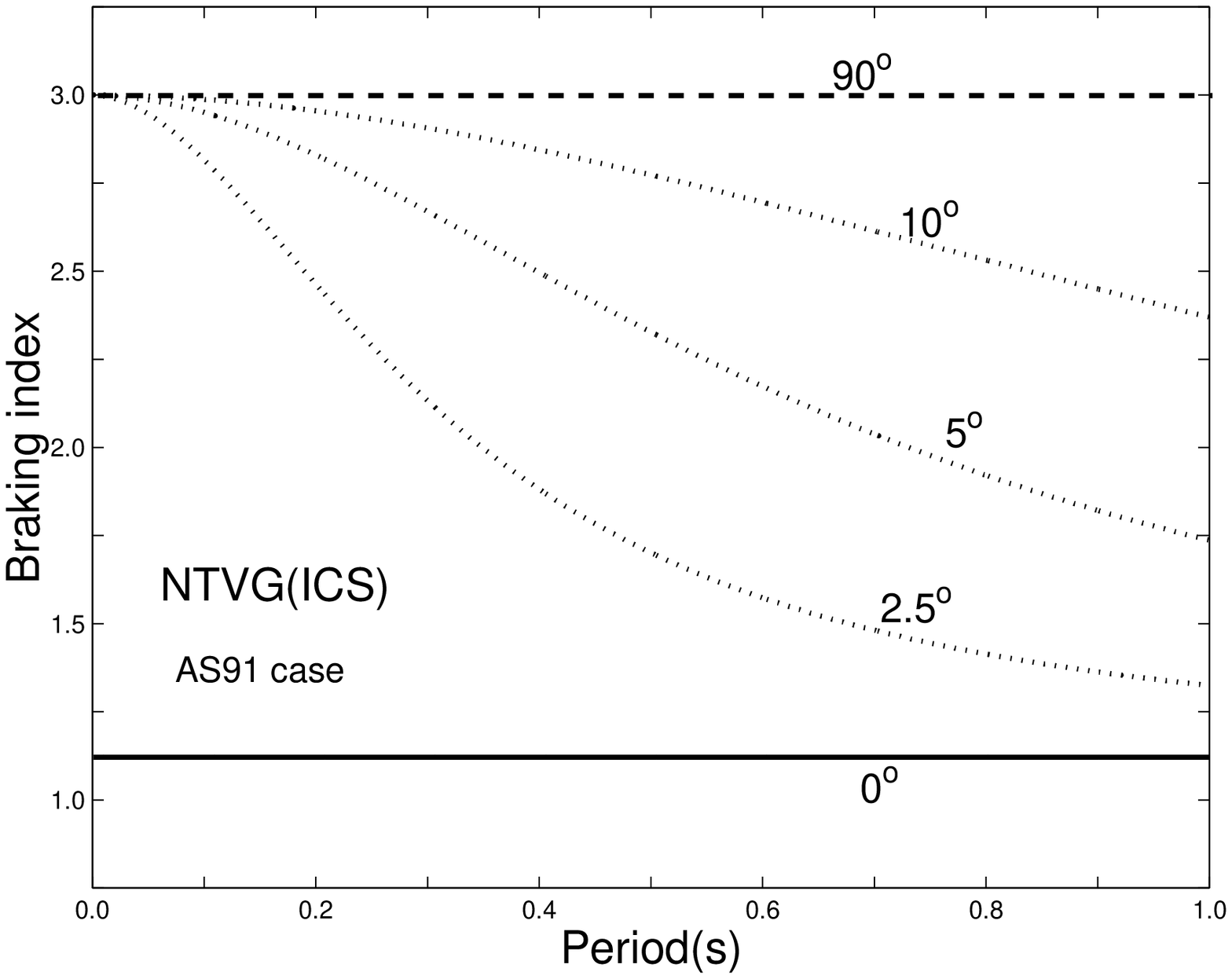}
  \end{minipage}%
  \begin{minipage}[t]{.5\textwidth}
    \centering
    \includegraphics[width=7cm]{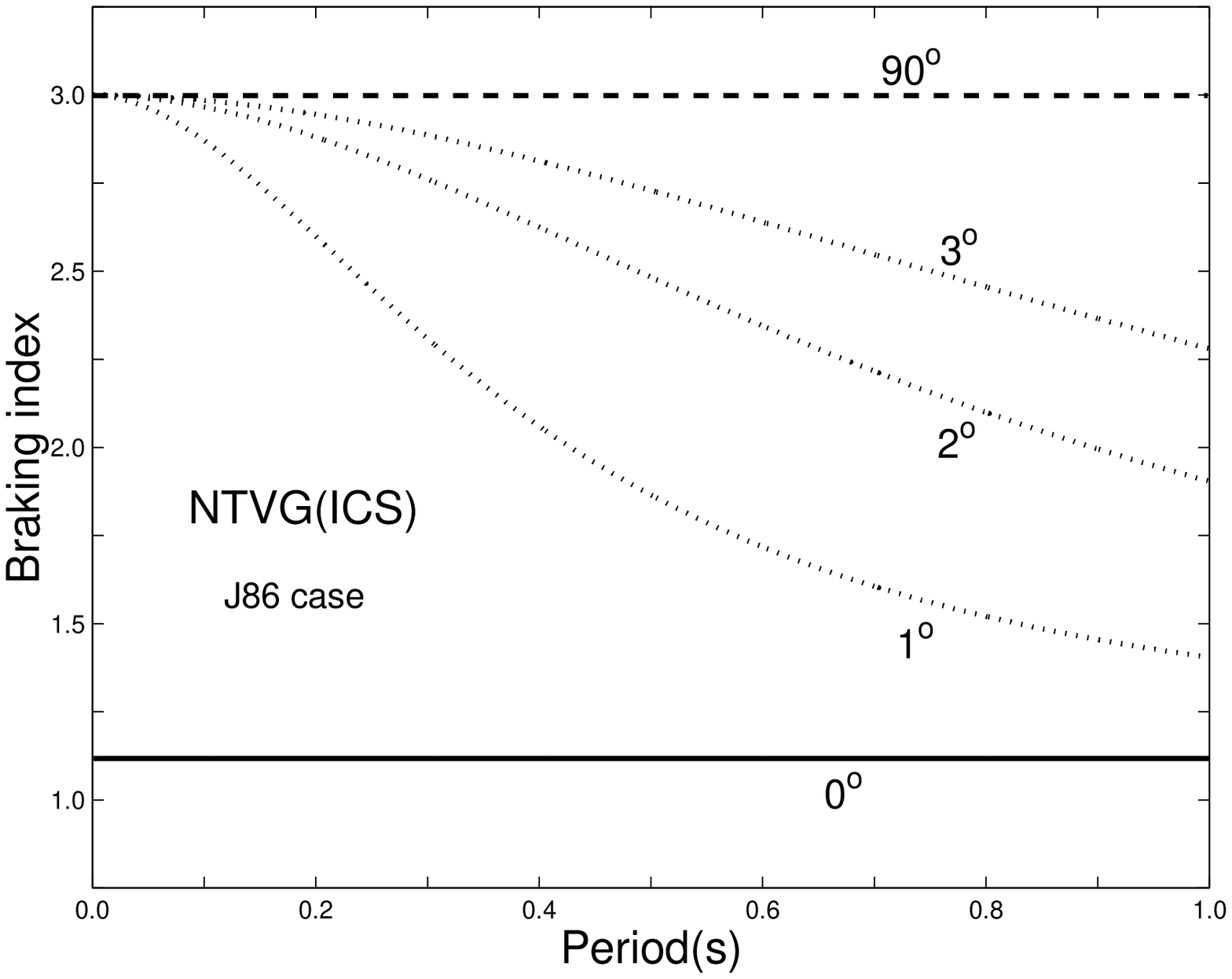}
  \end{minipage}%
     \caption{
Set of calculated braking indices, as functions of rotation
period, for NTVG models. Pulsars are assumed to have polar
magnetic field $B_{12}=10$ and radius $R=10^6\ \textrm{cm}$ here.
The curves presented are for pulsars with different inclination
angles, as indicated in the figure.
              }
\end{figure}


\clearpage
\begin{figure}
  \centering
    \includegraphics[width=12cm]{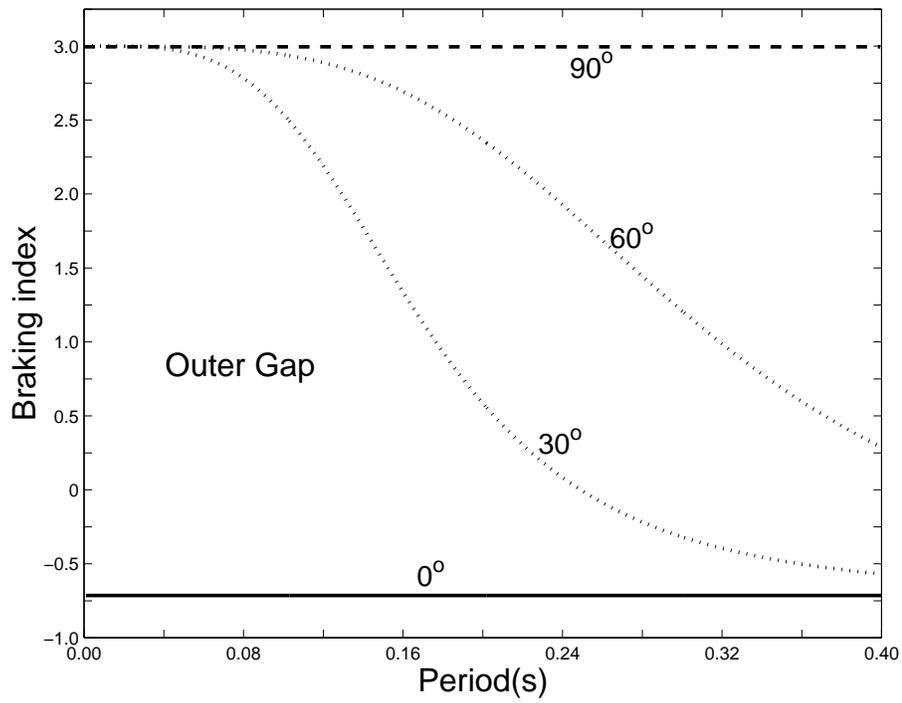}
    \caption{The braking index calculated in the outer gap model.
     See Fig.1 for other notes.}
\end{figure}


\clearpage

\begin{figure}[t]
  \centering
  \begin{minipage}[t]{.5\textwidth}
    \centering
    \includegraphics[width=7cm]{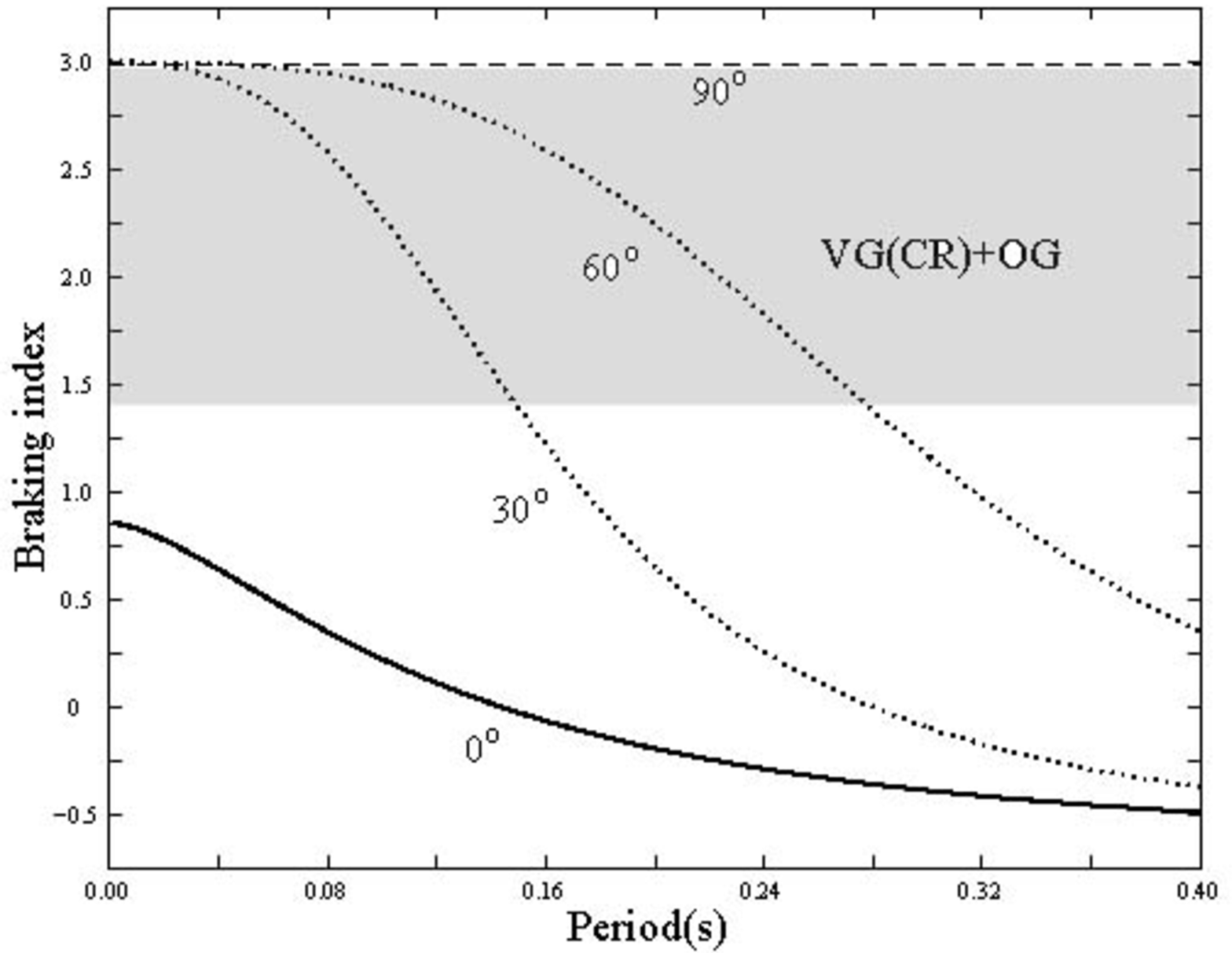}
  \end{minipage}%
  \begin{minipage}[t]{.5\textwidth}
    \centering
    \includegraphics[width=7cm]{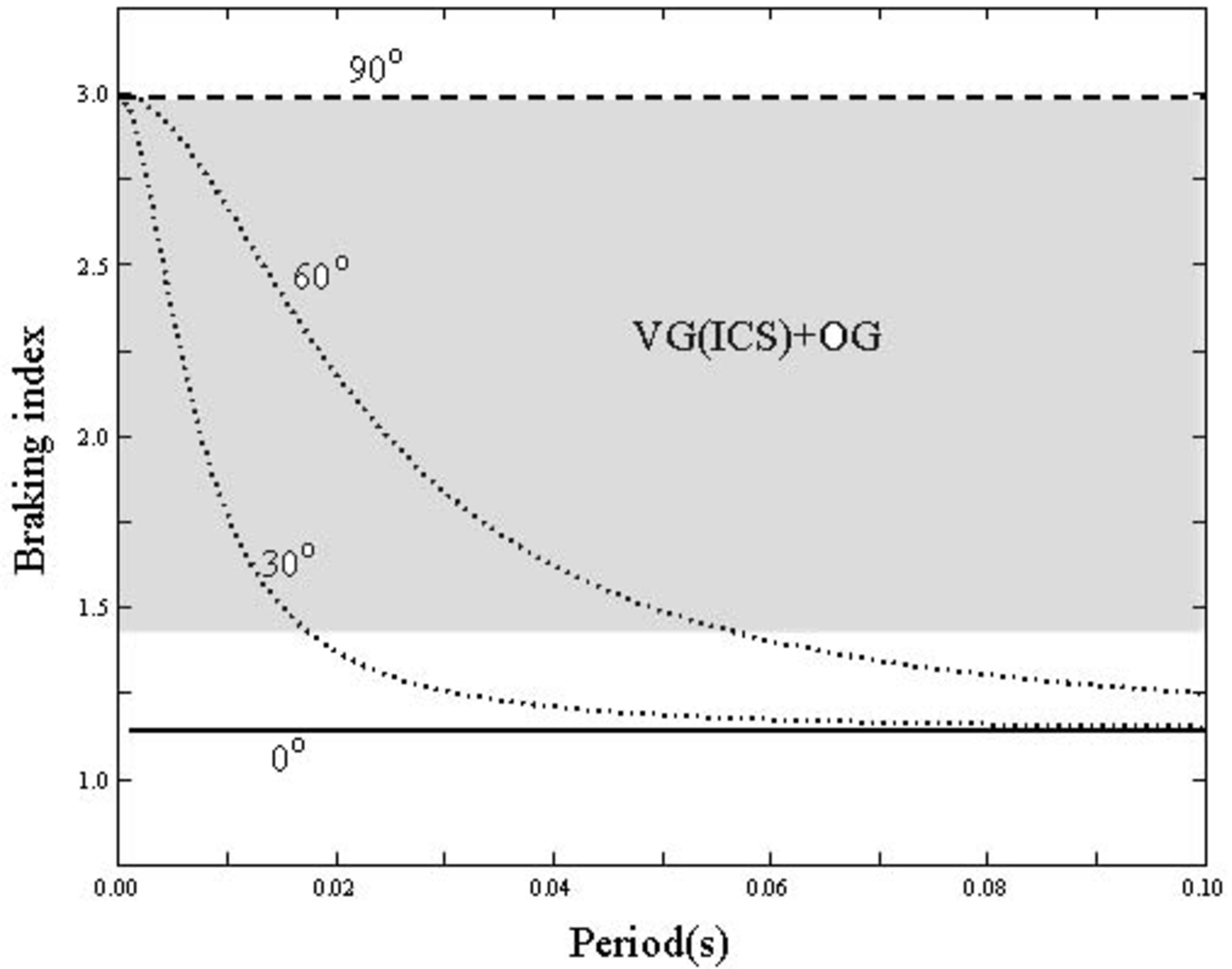}
  \end{minipage}%
\end{figure}

\begin{figure}
  \centering
  \begin{minipage}[t]{.5\textwidth}
    \centering
    \includegraphics[width=7cm]{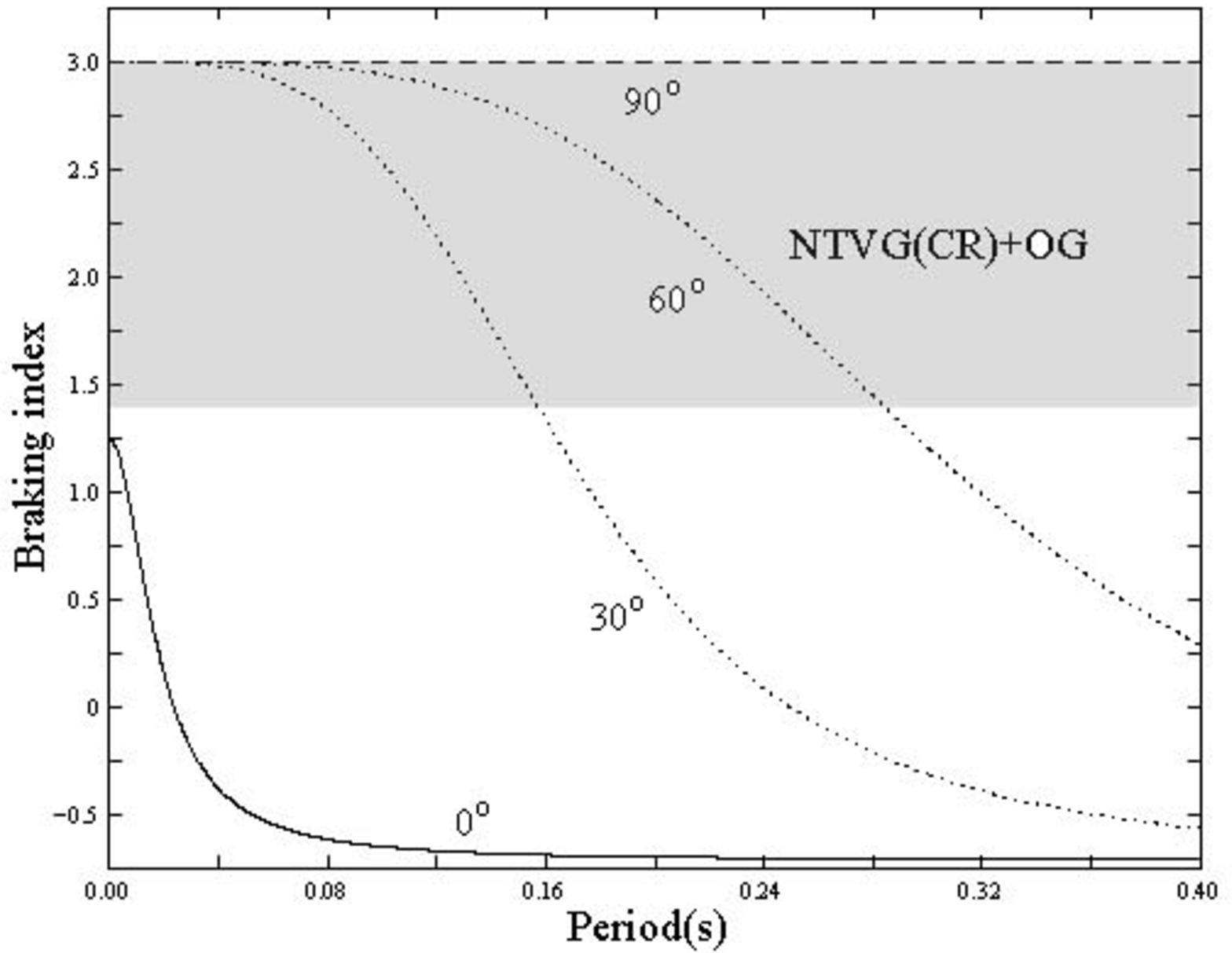}
  \end{minipage}%
  \begin{minipage}[t]{.5\textwidth}
    \centering
    \includegraphics[width=7cm]{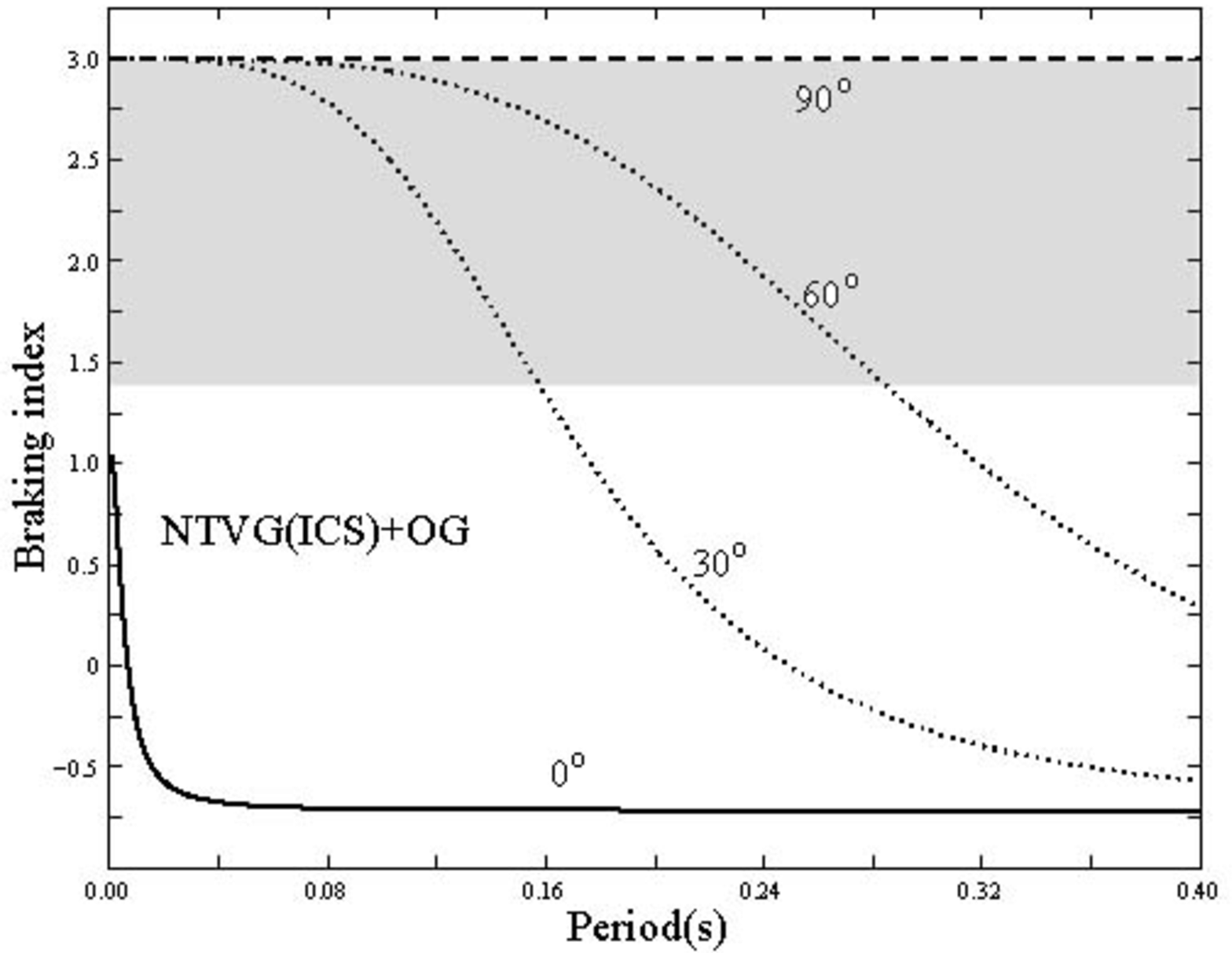}
  \end{minipage}%
\end{figure}

\begin{figure}
  \centering
  \begin{minipage}[t]{.5\textwidth}
    \centering
    \includegraphics[width=7cm]{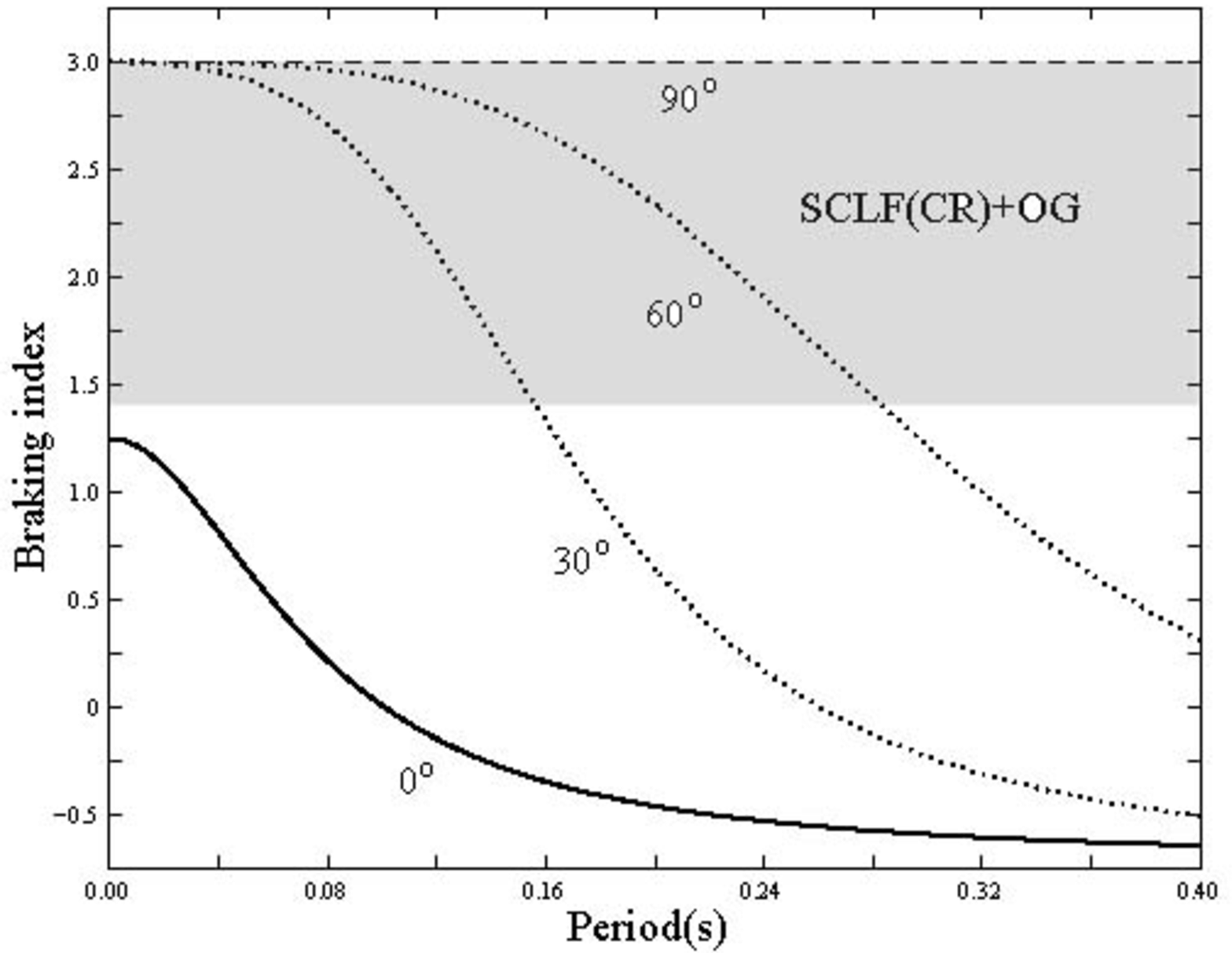}
    \caption{The braking indices calculated in the inner-outer gap models.
    The fields are chosen typically as $B_{12}=1$ except for the
    NTVG models for which we have $B_{12}=10$. The shadowed
    regions are for observed indices from 2.91 to 1.4.
     See Fig.1 for other notes.}
  \end{minipage}%
  \begin{minipage}[t]{.5\textwidth}
    \centering
    \includegraphics[width=7cm]{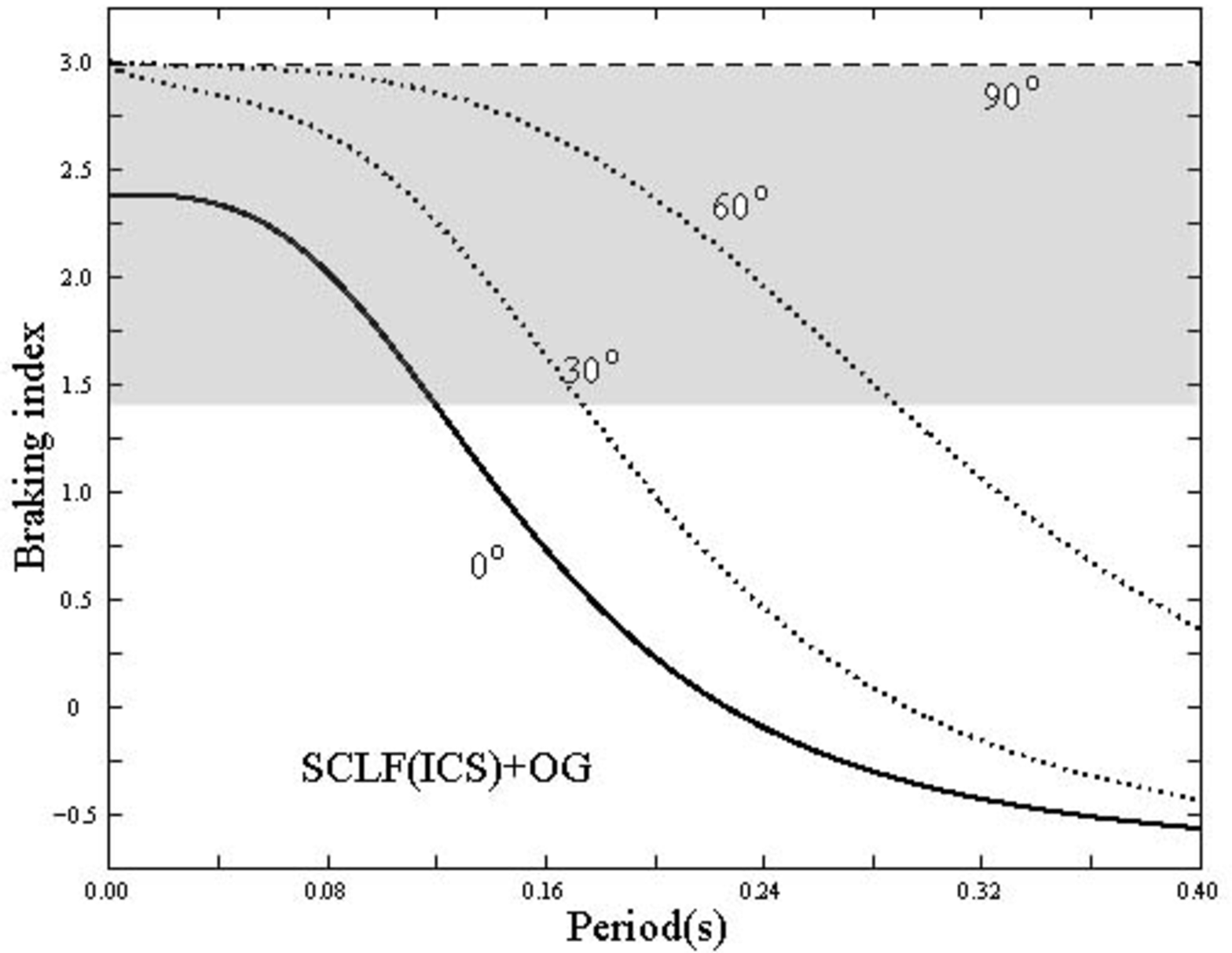}
  \end{minipage}%
\end{figure}

\end{document}